\documentstyle{fbssuppl}
\title{Relativistic corrections to the electric polarizability of the neutron}
\author{M. Bawin\instnr{1}, S. A. Coon\instnr{2} }
\instlist{Universit\'e de Li\`ege, Institut de Physique B.5, Sart Tilman, 
B-4000 Li\`ege 1, Belgium \and 
Physics Department, New Mexico State University, Las Cruces, NM 88003,
U.S.A.}

\begin{document}
\maketitle

\begin{abstract}
We demonstrate in a solvable model the connection between the
intrinsic  electric polarizability $\bar{\alpha}$ and the value
$\alpha_{Sch}$ obtained from neutron-atom scattering.
\end{abstract}

\section{Introduction}

According to the Low Energy Theorems of Compton scattering, the
electric $\bar{\alpha}$ and magnetic $\bar{\beta}$ polarizabilities are
fundamental quantities that characterize the neutron.  Given the lack
of free neutron targets and the much smaller Compton cross-section
(compared to the charged proton), the two current experimental
approaches are  low energy neutron-atom scattering~\cite{Jorg, Koester}
and quasi-free Compton scattering from the neutron bound in a
deuteron~\cite{Rose,WLS}.  The tool used to analyze the data in
neutron-atom scattering is the Foldy-Wouthuysen reduction of the Dirac
equation~\cite{Leeb}.  For a neutron of magnetic moment $\mu$ and mass
$m$, it has been  shown~\cite{BA97} that $\bar{\alpha}$ is {\em not}
the coefficient (labeled  $ \alpha_{Sch} $) of
$-{\textstyle\frac{1}{2}}  {\bf E}^2$ in this wave equation.  Instead, 
$\bar{\alpha} = \alpha_{Sch} + \mu^2/m$. 

The purpose of this communication is to illustrate these results by
solving exactly the the Foldy-Wouthuysen-Dirac equation for a neutron in
a constant ${\bf E}$ field.  Actually, this is how electric
polarizability is defined~\cite{Holstein}.  Our main result is the
following: the energy eigenvalues are independent of $\mu^2$ 
in the neutron rest frame defined by
\begin{equation} \langle m{\bf v} \rangle  =
 \langle {\bf p} - ({\bf E} \times \mbox{\boldmath $\mu$} )\rangle
  = 0 \,\,, 		
\label{velocity}
\end{equation}
where ${\bf p}$ is the canonical momentum,  ${\bf v}$ the velocity
operator, and $\mbox{\boldmath$\mu$} = \mu \mbox{\boldmath$\sigma$}$. 
Since the components of ${\bf v}$ implicitly include non-commuting spin
operators, the rest frame can only be defined in an average sense. 
Given this definition, the rest frame energy
eigenvalues do not include the $-\mu^2/m$ coefficient multiplying 
$-{\textstyle\frac{1}{2}}  {\bf E}^2$ present in the
Foldy-Wouthuysen-Dirac equation.
  Thus we confirm by an explicit calculation our earlier 
identification~\cite{BA97} $\bar{\alpha} = \alpha_{Sch} + \mu^2/m$
 which was of a somewhat formal nature.

\section{Calculation}

The Foldy-Wouthuysen-Dirac equation for a neutron of magnetic moment
$\mbox{\boldmath$\mu$} = \mu \mbox{\boldmath$\sigma$}$  and mass $m$ in
a constant external ${\bf E}$ field is~\cite{BA97}:

\begin{equation}    H \psi =
\left (\frac{{\bf p}^2}{2m} - \frac{{\bf p}}{m}\cdot ({\bf E}
   \times\mbox{\boldmath $\mu$} )
+ \frac{\mu^2 {\bf
E}^2}{2m} - {\textstyle\frac{1}{2}} \bar{\alpha} {\bf E}^2 \right)\psi =
\varepsilon \psi
\label{nonrel}
\end{equation}
where $\bar{\alpha}$ is the intrinsic neutron electric 
polarizability and $\varepsilon$ the energy eigenvalue.  Equation
(\ref{nonrel}) can be solved exactly to give:
\begin{equation}
  \left[ \varepsilon - \frac{ p^2}{2m} -\frac{\mu^2 E^2}{2m} +
  {\textstyle\frac{1}{2}} \bar{\alpha}  E^2 \right]
   = \frac{E \mu}{m} \lambda  \,\,\, ,
\end{equation}
where $\lambda$ is given by:
\begin{equation}
   {\bf p}\cdot (\mbox{\boldmath$\sigma$} \times {\bf E} ) \psi =
   E\lambda \psi \,\,\, ,
\end{equation}
and we have taken ${\bf E}$ to be along the $x$-axis, ie. :
\begin{equation}
	{\bf E} = (E,0,0)\,\,\,.
\end{equation}
Writing 
\begin{equation}
	{\bf p} = ( p_1, {\bf p}_{\perp})	
\end{equation}
one finds for (4):
\begin{equation}
	\lambda = \pm p_{\perp} \,\,, \hspace{0.25in} {\rm with}
	\hspace{0.25in} p_{\perp} = |{\bf p}_{\perp}|\,\,\,.
\end{equation}
With the aid of the explicit eigenstates of Eq. (4), one finds that the {\em
average} value of the  particle velocity operator
\begin{equation}  {\bf v} \equiv \frac{\partial H}{\partial {\bf p}} =
 \frac{1}{m} {\bf p} - ({\bf E} \times \mbox{\boldmath $\mu$} )		
\end{equation}
 for $\lambda = - p_{\perp}$ is
\begin{equation}
   \langle {\bf v}_{\perp} \rangle = \frac {{\bf p}_{\perp}}{m} 
    \left( 1 - \frac{\mu E} {p_{\perp}} \right)
\end{equation}
and
\begin{equation}
 \langle v_1 \rangle = v_1 = \frac {p_1} {m}
\end{equation}
Formulae (9) and (10) imply that
\begin{equation}
     \langle {\bf v} \rangle = 0 \hspace{0.5in}
      {\rm for} \hspace{0.5in}
     \mu E = p_{\perp}\,\,\,{\rm and}\,\, \, p_1 = 0\,\,\,  .
\end{equation}
This finally leads, from Eqs. (3), (11), and  $\lambda = - p_{\perp}$ to:
\begin{equation}
     \varepsilon = - {\textstyle\frac{1}{2}} \bar{\alpha} {\bf E}^2 
\end{equation}
Thus, in the neutron rest frame ($\langle {\bf v} \rangle = 0$), 
all terms quadratic in $\mu$ cancel exactly.
  This
implies, as stated in the Introduction, that the rest frame energy
eigenvalues do not include the $-\mu^2/m$ coefficient multiplying  
$-{\textstyle\frac{1}{2}}  {\bf E}^2$ present in Eq. (2).  It is
this Eq. (2), however, which
 is used to analyze neutron-atom scattering,  so that the quantity 
 $\mu^2/m$
should be {\em added} to the quoted polarizability result
($\alpha_{Sch}$)  of the experiment, thus
leading to an {\em increase} of the electric polarizability
$\bar{\alpha}$ (in the Compton sense), in complete
agreement with~\cite{BA97}.

\section{Discussion}

The explicit calculation just described sharpens the argument of
Ref.~\cite{BA97}.  In that earlier paper we asserted 
``that the rest frame of a neutron in an external 
electric field is defined by a vanishing value of the velocity operator, 
as confirmed by experimental measurements of the Aharonov-Casher
effect."  As we have seen in Section 2, the velocity {\em operator}
cannot vanish in a three dimensional geometry because it contains 
non-commuting spin operators,  and the neutron rest frame is defined by
the {\em expectation value} of the velocity operator.  The
observation~\cite{Opat} 
of the Aharonov-Casher effect was made in a two dimensional geometry (a
neutron diffracting around a line of electric charge) where the 
velocity {\em operator} does indeed vanish.

Finally, our result that the rest frame energy eigenvalues  do not
include the $-\mu^2/m$ coefficient multiplying  
$-{\textstyle\frac{1}{2}}  {\bf E}^2$ is also consistent with Foldy's
well known result.  Foldy solved exactly the problem of a 
structureless neutral Dirac particle with an anomalous magnetic  moment
$\mu$ in a homogeneous static electric field $\vec{E}$~\cite{Foldy}.
He  did find a   term quadratic in $\mu^2$ in the non-relativistic
limit of the energy eigenvalues ( implying a {\em negative}
polarizability of magnitude $\mu^2/m $  from this Dirac-Foldy term). 
Foldy's solution was obtained  in the frame ${\bf p} = 0$ which is not
the rest frame of the neutron.

\begin{acknowledge}
The work of M.B. was supported by the National Fund for Scientific
Research, Belgium and that of S.A.C. by NSF grant PHY-9722122.

\end{acknowledge}

\end{document}